 \documentclass[a4paper]{llncs}
\usepackage{svg}
\usepackage{color}
\usepackage{multirow}
\usepackage{rotating}
\usepackage{graphicx}
\usepackage[linesnumbered, vlined, ruled]{algorithm2e}
\usepackage{booktabs}
\usepackage{amssymb}
\usepackage{amsfonts}
\usepackage{amsmath}
\usepackage{soul}
\usepackage{etex}
\usepackage{extarrows}
\usepackage{relsize}
\usepackage{epstopdf}
\usepackage{url}
 \usepackage{hyperref} 
\hypersetup{ 
    colorlinks,%
    citecolor=black,%
    filecolor=black,%
    linkcolor=black,%
    urlcolor=black,%
    pdfproducer={nikos bikakis}, 
         pdftitle={rdf:SynopsViz -   A Framework for Hierarchical Linked Data Visual
 Exploration and Analysis},
             pdfauthor={Nikos Bikakis, Melina Skourla , George Papastefanatos},
         pdfsubject={rdf:SynopsViz -   A Framework for Hierarchical Linked Data Visual
 Exploration and Analysis},
pdfkeywords={rdf:SynopsViz} {SynopsViz} {HETree} {hierarchical visualization tree} {tree visualization structure} {hierarchical charts}  {hierarchical histogram} {hierarchical time line} {scalable visualization} {large data visualization}  {Visual analytics} {Semantic Web} {LOD} {RDF visualization} {Data exploration} {RDF Statistics} {RDF Charts} {Faceted search} {RDF Facets}
} 

\newcommand{\sviz}
{\textsf{rdf:SynopsViz }}

\newcommand{\eat}[1]{}

\newcommand{\stitle}[1]{\vspace{0.2cm}\noindent\textbf{#1}}

\begin{document}

\title{
 {rdf:SynopsViz}  --  A Framework for Hierarchical Linked Data Visual
 Exploration and Analysis \thanks{Further details regarding rdf:SynopsViz framework can be found at: Bikakis et al.: ``A Hierarchical Aggregation Framework for Efficient Multilevel Visual Exploration and Analysis", Semantic Web Journal  (2016).}\vspace{-2pt}}

 \author{
Nikos Bikakis$^{1,2}$ \hspace{3pt}
Melina Skourla$^{1}$\hspace{3pt} 
George Papastefanatos$^{2}$ 
}

\institute{
\vspace{-5pt} 
{$^{1}$National Technical University of Athens, Greece}\\
{$^{2}$IMIS, ATHENA Research Center, Greece}  \;
 }

\maketitle
 
\begin{abstract}
\vspace*{-15pt}
The purpose of data visualization is to offer intuitive ways for information
perception and manipulation, especially for non-expert users. The Web of Data
has realized the availability of a huge amount of datasets. However, the volume
and heterogeneity of available information make it difficult for humans 
to manually explore and analyse large  datasets.
In this paper, we present \textsf{rdf:SynopsViz}, a tool for hierarchical charting and
visual exploration of Linked Open Data (LOD). 
Hierarchical LOD exploration
is based on the creation of multiple levels of hierarchically related groups of
resources based on the values of one or more properties. 
The adopted hierarchical model 
provides    effective  information abstraction and summarization. 
Also, it allows efficient -on the fly- statistic computations, 
using aggregations over the hierarchy levels.
\vspace*{-5pt}
\end{abstract}


%


 \vspace*{-5pt}

\section{Introduction}
\label{sec:intro}
 \vspace*{-5pt}

The purpose of data visualization is to offer intuitive ways for information perception and manipulation 
that essentially amplify, especially for non-expert users, the overall cognitive
performance of information processing.
This is of great importance in the Web of Data, where the volume and heterogeneity of available information make 
difficult for humans to manually explore and analyse large  datasets. An important challenge is 
that visualization techniques must offer scalability and efficient processing
for     on the fly   visualization of large  datasets. They must also
employ appropriate data abstractions and aggregations for avoiding information overloading due to the size and diversity of the data presented to the user. 
Finally, they must be generic and provide uniform and intuitive visualization results across multiple domains.

In this work, we present \textsf{rdf:SynopsViz}, a framework for hierarchical charting and exploration
of Linked Open Data (LOD). Hierarchical LOD exploration realized through   the creation 
of multiple levels of hierarchically related groups of resources based on the values of one or more properties.
For example, a numerical group, characterized by a numerical range, comprises all resources 
with a property value within the range of this group. Hierarchical browsing can address the problem of 
information overloading as it provides information abstraction and summarization \cite{EF10}. It can also offer rich 
insights on the underlying data when combined with rich statistical information on the groups and their contents. 

The key features of \sviz  framework are summarized as follows:
(1) It adopts a \textit{hierarchical model} for RDF data visualization, browsing and analysis. 
(2) It offers \textit{automatic} on-the-fly  hierarchy construction based on data distribution, 
as well as \textit{user-defined} hierarchy construction based on user's preferences.
(3) Provides  \textit{faceted}    browsing and filtering  over  classes and properties.
(4) Integrates \textit{statistics with visualization};
visualizations have been enriched with useful statistics and data information. 
(5) Offers several visualizations techniques  (e.g., timeline, chart, treemap).
(6) Provides a large number of dataset's \textit{statistics} regarding the:
data-level (e.g., number of sameAs triples),  
schema-level (e.g., most common classes/properties), 
and structure level (e.g., entities with the larger in-degree).
(7) Provides numerous  \textit{metadata}  
related to the dataset: licensing, provenance, linking, availability, undesirability, etc.
The latter  are useful for assessing   data quality \cite{ZRMP+13}.
 

\begin{figure}[b] 
 \vspace{-6mm}
\centering
\includegraphics[scale=1]{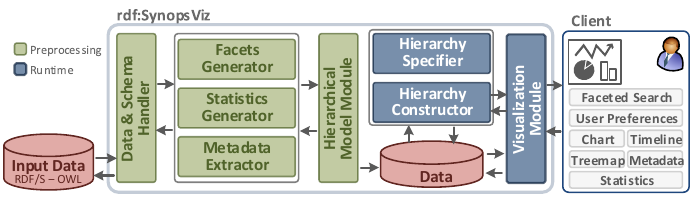}
\vspace{-4mm}
\caption{System Architecture}
\label{fig:arch}
\end{figure}


\section{Framework Overview}
\label{sec:over}

The architecture of \sviz is presented in Figure~\ref{fig:arch}.
Our scenario involves three main parts: the Client GUI, the  \sviz framework,
and the input data. The \textit{Client} part, corresponds to the framework's
front-end offering several functionalities to the end-users (e.g., statistical
analysis, facet search, etc.).
\sviz consumes RDF data as \textit{Input data}; optionally, OWL-RDF/S
vocabularies/ontologies describing the input data can be loaded.
 Next,  we describe the basic components of the \sviz  framework. 

In the preprocessing phase,  the \textit{Data and Schema Handler}
parses the input data and inferes schema information (e.g., properties domain(s)/range(s), class/property  hierarchy, type of
instances, type of properties, etc.).
\textit{Facets Generator} generates class and property facets over input data. 
\textit{Statistics Generator} computes several statistics regarding the
schema, instances and graph structure of the input dataset, such as the number
of different types of classes and properties, or the number of sameAs triples,
 or finally the average in/out degree of the RDF graph, respectively.
\textit{Metadata Extractor} collects dataset metadata which can be used for data quality assessment.
\textit{Hierarchical Model Module}
adopts our hierarchy model and stores the initial data enriched with the information computed during the preprocessing phase.

During runtime the following components are involved.
\textit{Hierarchy Specifier} is responsible for managing the
configuration parameters of our hierarchy model, e.g., the number of
hierarchy levels, the number of nodes per level, and providing this
information to the Hierarchy Constructor. 
\textit{Hierarchy Constructor}
implements the hierarchy model. 
Based on the selected facets, and the hierarchy configuration: 
 it determines the hierarchy of groups and the contained triples, and computes the 
statistics about their contents (e.g., range, variance, mean, number of triples contained, etc.).
\textit{Visualization Module} allows the interaction between the user and the framework, 
allowing several operations (e.g, navigation, filtering, hierarchy specification) over the visualized data.


\section{Implementation \& Demonstration Outline}
\stitle{Implementation.}
\sviz is implemented on top of several open source tools and libraries. 
Regarding visualization libraries, we use
Highcharts\footnote{\href{http://www.highcharts.com}{www.highcharts.com}}, for
the area and timeline charts. and Google
Charts\footnote{\href{https://developers.google.com/chart}{developers.google.com/chart}}  for treemap and pie charts.
Additionally, it uses Jena
framework\footnote{\href{http://jena.apache.org}{jena.apache.org}} for RDF data handing and Jena TDB for RDF storing.

 
The web-based prototype of \sviz is
available at \url{http://synopsviz.imis.athena-innovation.gr}. Also a video demonstrating the
scenario presented below is available at 
\url{http://youtu.be/8v-He1U4oxs}.
%

\stitle{Demonstration scenario.}
First, the attenders will be able to
 select a dataset from a number of offered real-word datasets 
(e.g., dbpedia,  Eurostat, World Bank, U.S. Census, etc.) or upload their own. 
Then, for the selected dataset, the attendees are able to examine several of the dataset's  \textit{metadata}, 
 and explore several datasets's \textit{statistics}.

Using the facets panel, the attenders are able to navigate and filter data based on classes, numeric and date properties.
In addition, through facets navigation several information about the classes and properties 
(e.g.,  number of instances, domain(s), range(s), IRI, etc.) are provided to the users through the UI.

The attenders are able to navigate over data by considering  properties' values. 
Particularly, area charts and timeline-area charts are used to 
visualize the resources considering  the user's selected properties. 
Classes' facets can also be used to filter the visualized data.
Initially, the top level of the hierarchy is
presented providing an overview of the data, organized into top-level groups;
the user can interactively zoom in and out the group of interest, up to
the actual values of the raw input data. At the same time, statistical information concerning the hierarchy groups as well as their 
contents (e.g., mean value, variance, sample data, etc.) are presented.

In addition, the attenders are able to navigate over data, through class hierarchy. 
Selecting one or more classes, the attenders can interactively navigate over the class hierarchy using treemaps. 
In \sviz the treemap visualization has been enriched with schema and statistical information.
For each class, schema metadata (e.g., number of instances,
 subclasses, datatype/object  properties) and statistical information
  (e.g., the cardinality of each property,  min, max value for datatype properties' ranges, etc.) are provided. 

Finally, the attenders can interactively modify the hierarchy specifications. 
Particularly, they are able to increase or decrease the level of abstraction/detail presented, by modifying modifying both the number of hierarchy levels, 
and number of nodes per level.

\eat{ 
\begin{figure}[t]
 \centering
 \vspace{-18mm}
\includegraphics[scale=0.25]{figures/ui}
\vspace{-2mm}
\caption{Web User Interface}
\vspace{-8mm}
\label{fig:UI}
\end{figure} }

 \section{Related Work} 

A large number of works studying issues related to RDF or LOD
visualization and analysis have been proposed in the literature
\cite{DR11,BAG12,DRP11,AKSH13}.
Additionally, numerous tools offering  RDF or Linked Open Data visualization have been developed, e.g., 
 \textit{Sgvizler} \cite{S12},
 \textit{LODWheel} \cite{SDN11},
\textit{Payola} \cite{KHN13},
\textit{CubeViz} \cite{SMB+12},
\textit{KC-Viz} \cite{MMP+11},
\textit{RelFinde}\footnote{\href{http://www.visualdataweb.org/relfinder.php}{www.visualdataweb.org/relfinder.php}},
\textit{Welkin}\footnote{\href{http://simile.mit.edu/welkin}{simile.mit.edu/welkin}}, 
\textit{IsaViz}\footnote{\href{http://www.w3.org/2001/11/IsaViz}{www.w3.org/2001/11/IsaViz}},
\textit{RDF-Gravity}\footnote{\href{http://semweb.salzburgresearch.at/apps/rdf-gravity}{semweb.salzburgresearch.at/apps/rdf-gravity}}, etc.

In the context of  RDF and Linked Open Data statistics,  
\textit{RDFStats} \cite{LW09} calculates statistical information about RDF datasets. 
\textit{LODstats} \cite{ADML12} is  an extensible framework, offering scalable statistical analysis of Linked Open Data datasets.

Regarding the quality assessment issues,
\cite{ZRMP+13} studies the criteria which can be used in   Linked Data quality assessment. 
  \cite{LW09} review millions of  RDF documents to analyse Linked Data conformance.
Finally, several frameworks for the quality assessment in the Web of Data, have been proposed 
  \textit{LINK-QA} \cite{GGSL12}, \textit{Sieve} \cite{MMB12}, \textit{WIQA}
  \cite{BC09}.
  In contrast to existing approaches, we provide hierarchical RDF
  data visualization enriched with data statistics. The hierarchical model
  solves the visualization overload issues, offering efficient, on the fly statistical computations over hierarchy levels.
  Finally, due to hierarchical model our tool can efficiently handle and analyse very large datasets.

 
 \section{Conclusions}

In this paper we have presented \textsf{rdf:SynopsViz},
a framework for hierarchical charting and exploration of Linked Open Data. 
The hierarchical model adopted by our framework can 
address the problem of information overloading, offering an effective mechanism for information abstraction and summarization.
Additionally, the adopted model allows the efficient statistic computations, 
using aggregations over the hierarchy levels.

Some future extensions of our tool include the application of more sophisticated
filtering techniques (e.g., SPARQL-enabled browsing over the data), as well as the
addition of more visual techniques and libraries.

\begin{small}
 \stitle{Acknowledgement.} This research has been co-financed by the European Union (European Social Fund - ESF) and Greek national funds through the Operational Program "Education and Lifelong Learning" of the National Strategic Reference Framework (NSRF) - Research Funding Program: THALIS and KRIPIS - Investing in knowledge society through the European Social Fund. 
\end{small}

\end{document}